\documentclass[pre,amssymb,superscriptaddress,showpacs,floatfix,a4paper,twocolumn]{revtex4}

\usepackage{graphicx}
\usepackage{dcolumn}
\usepackage{bm}

\begin{document}

\title{Backbone structure of the Edwards-Anderson spin-glass model}

\author{F. Rom\'a}
\affiliation{Departamento de F\'{\i}sica, INFAP, CONICET, Universidad Nacional de
San Luis, Chacabuco 917, D5700BWS San Luis, Argentina}
\author{S. Risau-Gusman}
\affiliation{Centro At\'omico Bariloche, CONICET, San Carlos de
Bariloche, R8402AGP R\'{\i}o Negro, Argentina }

\date{\today}

\begin{abstract}
We study the ground-state spatial heterogeneities of the Edwards-Anderson spin-glass model with 
both bimodal and Gaussian bond distributions.  We characterize these heterogeneities by using a 
general definition of bond rigidity, which allows us to classify the bonds of the system into two 
sets, the backbone and its complement, with very different properties. This generalizes to continuous 
distributions of bonds the well known definition of a backbone for discrete bond distributions. By extensive 
numerical simulations we find that the topological structure of the backbone for a given lattice 
dimensionality is very similar for both discrete and continuous bond distributions. We then 
analyze how these heterogeneities influence the equilibrium properties at finite 
temperature and we discuss the possibility that a suitable backbone picture can be relevant to 
describe spin-glass phenomena. 
\end{abstract}


\maketitle

\section{Introduction \label{Intro}}

Spin glasses are the paradigm of systems exhibiting both quenched disorder 
and frustration \cite{Binder1986}. In such magnetic materials, static and 
dynamical behaviors are far from being completely understood. Even though 
in the absence of an external magnetic field, experiments, theory, and numerical 
simulations agree on the existence of a phase transition at a finite temperature, 
there is still controversy regarding the true nature of the low temperature phase. 
In this matter, two different theories have dominated the field for many years. One of them uses 
the concept of replica symmetry breaking \cite{Parisi1983} to go beyond mean-field 
methods, and predicts that spin glasses have a nontrivial phase space broken in many 
ergodic components and with an ultrametric topology. Unlike this complex scenario, 
the phenomenological {\em droplet picture} \cite{Fisher1986} postulates a simpler 
structure for the phase space, with only two pure states related to each other 
by an up-down symmetry.  Most of the experimental and numerical results have been 
interpreted in the light of these two theories.  As the controversy persists, 
there have been other attempts to explain, within a single framework, many of 
the results reported in the literature \cite{Krzakala2000,Palassini2000,Marinari2001,Middleton2001,Katzgraber2001,NewmanStein,Lamarcq2002}.  

Recently a new approach \cite{Roma2006,Roma2007a,Roma2007b,Roma2010a,Roma2010b,Rubio2010a,Rubio2010b} 
has been put forward that can provide a new way to reinterpret some of the 
numerical and theoretical data in the literature. Based on the same spirit as 
the droplet picture, which focuses on the ground state (GS) and its excitations, 
in this approach the spatial heterogeneities of the GS play a fundamental 
role in describing the low-temperature behavior of the system. In the Edwards-Anderson 
$\pm J$ model \cite{EA}, which has a degenerate GS, these heterogeneities are characterized 
by the {\em backbone}, defined as the union of the {\em rigid lattice} and the {\em solidary spins}. 
This rigid lattice \cite{Barahona1982} is the set of bonds (called {\em rigid bonds}) which do not change its condition 
(satisfied or frustrated) in all the configurations of the GS. The remaining ones, called {\em flexible bonds}, 
form the {\em flexible lattice}. The solidary spins are the spins which maintain their 
relative orientation in all configurations of the GS (the remaining spins are called {\em nonsolidary} spins).

It has been shown \cite{Roma2006,Roma2007a,Roma2007b,Roma2010a,Roma2010b,Rubio2010a,Rubio2010b} 
that the backbone structure is closely linked to the static and the dynamical behavior 
of the Edwards-Anderson $\pm J$ model.  For instance, in the out-of-equilibrium dynamics the 
mean flipping time probability distribution function has two main peaks corresponding 
to fast and slow degrees of freedom \cite{Ricci-Tersenghi2000}, and in Refs. \onlinecite{Roma2006,Roma2010a} 
it was shown that these peaks are directly related to the nonsolidary and the solidary spins, respectively. 
In addition, for long simulation times the clusters of nonsolidary spins satisfy the fluctuation-dissipation 
theorem, whereas the solidary spins violate this relation, even below the critical temperature \cite{Roma2007b}.  
Thus, the backbone and its complement can be associated with a spin-glass phase and a paramagnetic 
phase, respectively.

These numerical results suggest that a suitable {\em backbone picture} can be relevant to 
describe the physics of spin glasses. However, in order to build a more comprehensive theory 
we need to define the backbone in other disordered and frustrated systems \cite{Roma2010b}. 
Here we generalize the concept of bond {\em rigidity}, which allows us to speak of the 
{\em rigid structure} of each sample, to cover the cases of Ising models with nondegenerate GS \cite{Roma2010b}. 
In the next sections we show that using this idea it is possible to define a backbone having 
the same physical and topological properties as in spin-glass systems with degenerate GS.

The paper is structured as follows.  In Sec. \ref{ModRig} we present the Edwards-Anderson model 
for bimodal and Gaussian distributions of bonds and we define the concepts of rigidity and rigid structure.  
In Sec.\ref{NumRes} a numerical study of the topology of this structure, including a percolation 
analysis, is presented, as well as a study of some physical properties. 
In the last section we discuss our results and some conclusions are drawn. 
In particular, we address the important issue of {\em temperature-chaos}, 
an effect present in spin glasses and common to both, the droplet \cite{McKay1982,Bray1987} and 
the mean-field pictures \cite{Rizzo2003,Parisi2010}.  This phenomenon refers to the fragility of 
the equilibrium state on small temperature changes.  Recent simulation studies \cite{Katzgraber2007,Fernandez2013} 
of the Edwards-Anderson model show that, even for small systems, some rare samples are significantly 
affected by temperature chaos.  In this context, because the phenomenological backbone picture 
relies on simulations of very small lattices, it could be argued that our findings are not 
relevant to finite temperature in the thermodynamic limit.  We argue that, if the concept of bond 
rigidity is interpreted in terms of ``effective interactions,'' it is reasonable to expect that the 
GS structures we are considering are linked to the finite-temperature behavior of the system.

\section{\label{ModRig} The Edwards-Anderson model and the GS rigid structure}

We start by considering the Hamiltonian of the Edwards-Anderson spin-glass model \cite{EA}:
\begin{equation}
\mathcal{H} = - \sum_{( i,j )} J_{ij} \sigma_{i} \sigma_{j}, \label{ham}
\end{equation}
where the sum runs over the nearest-neighbor sites of either a two-dimensional (2D) square or a three-dimensional 
(3D) cubic lattice of linear dimension $L$ and $\sigma_i = \pm 1$ are $N$ Ising spin variables.  The 
coupling constants $J_{ij}$ are independent random variables chosen from a bimodal distribution,
\begin{equation}
D_{\mathrm{B}}(J_{ij})=\frac{1}{2} \left[\delta(J_{ij}-1)+\delta(J_{ij}+1)\right],   \label{DistB}
\end{equation}
or a Gaussian distribution,
\begin{equation}
D_{\mathrm{G}}(J_{ij})= \frac{1}{\sqrt{2 \pi}} \exp(- J_{ij}^2/2), \label{DistG}
\end{equation}
\noindent for which the mean value is zero and the variance is $1$.  These are the most often used
bond distributions.  Hereafter, the versions of the Edwards-Anderson model where interactions 
are drawn from Eqs. (\ref{DistB}) and (\ref{DistG}) are called EAB and EAG, respectively.  The samples 
analyzed were generated with both periodic-free boundary 
conditions (pfbc) and periodic-periodic boundary conditions (ppbc) in 2D, whereas in 3D only periodic boundary 
conditions in all directions were used.

As discussed in the Introduction, for the EAB model, and for any Ising system with a 
degenerate GS, it is possible to define a rigid lattice and a set of solidary spins, which can in turn 
be used to define a backbone. These definitions, however, make use of the degeneracy of the GS and thus 
cannot be used in a system with a single GS. To understand how to generalize the definitions for such 
systems, following Ref. \onlinecite{Roma2010b} we consider the Edwards-Anderson model with a continuous 
distribution of bonds that consists of the superposition of two Gaussian functions of width (variance) 
$\epsilon$ centered at $J=\pm 1$.  We call this the EAB-$\epsilon$ model.  If $\epsilon$ is small enough, 
the physical properties of each sample of the EAB-$\epsilon$ model will be very close to the ones of its EAB 
``companion sample,'' obtained by replacing each ferromagnetic (antiferromagnetic) bond $J_{ij}$ by $+1$($-1$).  
In particular, the GS configurations of the companion sample correspond to the lowest excited states of 
the EAB-$\epsilon$ sample. It seems reasonable to define the backbone of this system as the same set of 
spins and bonds as in the companion sample. Thus, the spins and bonds of this backbone have the same 
orientations and condition, respectively, in the lowest excited states of the EAB-$\epsilon$ sample.
This example suggests that, to generalize the definition of backbone to an arbitrary model, 
it is necessary to consider not only the GS but also the low-excitation levels.  In particular the {\em rigidity} 
of each bond should be a parameter taking a continuum of values, instead of only two (rigid-flexible) 
as in the EAB model.  A definition was proposed in Ref. \onlinecite{Roma2010b}, and is as follows. 
Consider a sample of the Edwards-Anderson model with an arbitrary bond distribution (discrete or continuous).  
For each bond $J_{ij}$ we define its rigidity as $r_{ij}=U_{ij}-U$, where $U$ is the GS energy of the sample 
and $U_{ij}$ is the lowest energy for which the bond $J_{ij}$ is frustrated (satisfied) 
if it is satisfied (frustrated) in the GS.  The {\em rigid structure} (RS) of a sample is then defined as a 
lattice where each bond $J_{ij}$ has been replaced by its rigidity $r_{ij}$.   As shown in the following sections, 
the set of the most rigid bonds of the RS seems to behave as a backbone of the EAG model. 

\section{\label{NumRes} Numerical results}

To find the RS of each sample, for both the EAB and the EAG, the method we use is very similar to that used to find 
the rigid lattice of the EAB model \cite{Roma2010b,Ramirez2004}.  Assuming that one has an algorithm for 
obtaining GS configurations, a scheme of the procedure is as follows: 
\begin{enumerate}
\item{A GS configuration, $C$, is found and its energy $U$ is stored.}

\item{A bond $J_{ij}$ is chosen.}

\item{One of the spins joined by the bond $J_{ij}$, i.e. either $\sigma_{i}$ or $\sigma_{j}$, is flipped. 
This changes the ``condition" of the bond from satisfied to frustrated or viceversa.}

\item{The orientations of the spins $\sigma_{i}$ and $\sigma_{j}$ are frozen.}

\item{For this ``constrained'' system a GS configuration $C^*$ is found, and its energy $U_{ij}$ is stored.}

\item{The rigidity of the selected bond is calculated as $r_{ij}=U_{ij}-U$.}

\item{The process is repeated from step (2) until the rigidity of every bond has been calculated.}

\end{enumerate}
Note that the number of GSs that have to be calculated to obtain the RS is equal to the number of bonds.
Furthermore, note that only the energy of the GS is really necessary for this procedure to work. 

In the procedure above it is assumed that the algorithm for finding the GS is deterministic.  But for 
some systems, if the sample size is not very small, only probabilistic algorithms are 
available, i.e., algorithms whose output is a GS configuration with a probability smaller than $1$. In 
this case, the only modification to the previous procedure is that, in steps (1) and (5), we perform $n$ 
independent runs of the probabilistic algorithm [evidently, if in step (5) we obtain $U_{ij}=U$, no further 
runs are performed], in order to find a reliable GS configuration.  

For lattices with ppbc we use a parallel tempering Monte Carlo algorithm \cite{Geyer1991,Hukushima1996}. 
It has recently been shown that this technique is a powerful heuristic method for reaching the GS of the 
EAB and the EAG models in both 2D and 3D lattices \cite{Roma2009}.  As many independent runs of this 
algorithm are needed to determine the RS of each sample, we have only been able to study 
lattice sizes of the EAB (EAG) model up to $L=22$ ($L=16$) in 2D and $L=10$ ($L=8$) in 3D. For simplicity, 
in all cases we used $m=20$ replicas of the system, and the highest and lowest temperatures 
were set at $T_1=1.6$ and $T_m=0.1$.  For each model with ppbc, Tables~\ref{table2D} and \ref{table3D} 
list the remaining parameters used in our simulations for the different lattice sizes: the total number 
of Monte Carlo sweeps $N_\mathrm{sw}$, the number of samples $N_\mathrm{sa}$, and the parameter $n$.  

\begin{table}[t]
\caption{\label{table2D} Simulations parameters for the 2D EAB and EAG models with ppbc (see text).}
\begin{tabular}{cccccccc}
\hline
\hline
       &      &  EAB              &                 &        &      & EAG             &                 \\
\hline
$L$    & $n$  & $N_\mathrm{sw}$   & $N_\mathrm{sa}$        & $L$    & $n$  & $N_\mathrm{sw}$      & $N_\mathrm{sa}$           \\ 
\hline
$6-16$ & $8$  & $2\times 10^4$    & $10^4$          & $6-12$ & $8$  & $10^5$          & $10^4$           \\ 
$18$   & $10$ & $10^5$            & $6 \times 10^3$ & $14$   & $12$ & $2 \times 10^5$ & $5 \times 10^3$  \\ 
$20$   & $10$ & $10^5$            & $3 \times 10^3$ & $16$   & $10$ & $6 \times 10^5$ & $3 \times 10^3$  \\ 
$22$   & $10$ & $2.5 \times 10^5$ & $2 \times 10^3$ &        &      &                 &                  \\ 
\hline
\hline
\end{tabular}
\end{table}

\begin{table}[t]
\caption{\label{table3D} Simulations parameters for the 3D EAB and EAG models (see text).}
\begin{tabular}{cccccccc}
\hline
\hline
       &      &  EAB              &             &        &      & EAG             &                  \\
\hline
$L$    & $n$  & $N_\mathrm{sw}$   & $N_\mathrm{sa}$    & $L$    & $n$  & $N_\mathrm{sw}$        & $N_\mathrm{sa}$         \\ 
\hline
$4$    & $10$ & $5\times 10^3$    & $10^4$      & $4$    & $10$ & $2\times 10^4$  & $10^4$           \\ 
$6$    & $10$ & $2\times 10^4$    & $10^4$      & $5$    & $10$ & $5\times 10^4$  & $10^4$           \\ 
$8$    & $10$ & $7\times 10^5$    & $10^3$      & $6$    & $10$ & $3\times 10^5$  & $10^4$           \\ 
$10$   & $40$ & $2\times 10^6$    & $10^2$      & $7$    & $12$ & $5\times 10^5$  & $3\times 10^3$   \\ 
       &      &                   &             & $8$    & $12$ & $10^6$          & $10^3$           \\ 
\hline
\hline
\end{tabular}
\end{table}

For planar lattices it is well known that the problem of finding GS configurations 
can be mapped to a minimum-weighted perfect matching problem, which can be solved exactly in polynomial 
time (i.e, in time proportional to some power of $L$) \cite{Hartmannlibro}. Then, to study 2D samples with 
pfbc, we have used one implementation of the Blossom algorithm \cite{Blossom} which has allowed us to 
obtain the RS of larger systems sizes.  Tables~\ref{table2Dpfbc} shows the corresponding parameters.
The largest sample size that we have studied is $L=60$, which is smaller than the sizes studied in 
Ref. \onlinecite{Roma2010b}, because to determine the RS many more GSs are needed than for the rigid lattice.

\begin{table}[t]
\caption{\label{table2Dpfbc} Parameters for the 2D EAB and EAG models with pfbc (see text).}
\begin{tabular}{cccc}
\hline
\hline
          &        EAB       &           &     EAG                             \\
\hline
$L$       &  $N_\mathrm{sa}$ & $L$       &  $N_\mathrm{sa}$           \\ 
\hline
$10-30$   &  $10^4$          & $10-30$   &  $10^4$           \\ 
$40$      &  $6\times 10^3$  & $40$      &  $2\times 10^3$           \\ 
$50$      &  $2\times 10^3$  & $50$      &  $10^3$           \\ 
$60$      &  $3\times 10^2$  & $60$      &  $5\times 10^2$           \\ 
\hline
\hline
\end{tabular}
\end{table}

\subsection{Rigidity distribution}

\begin{figure}[t!]
\includegraphics[width=7cm,clip=true]{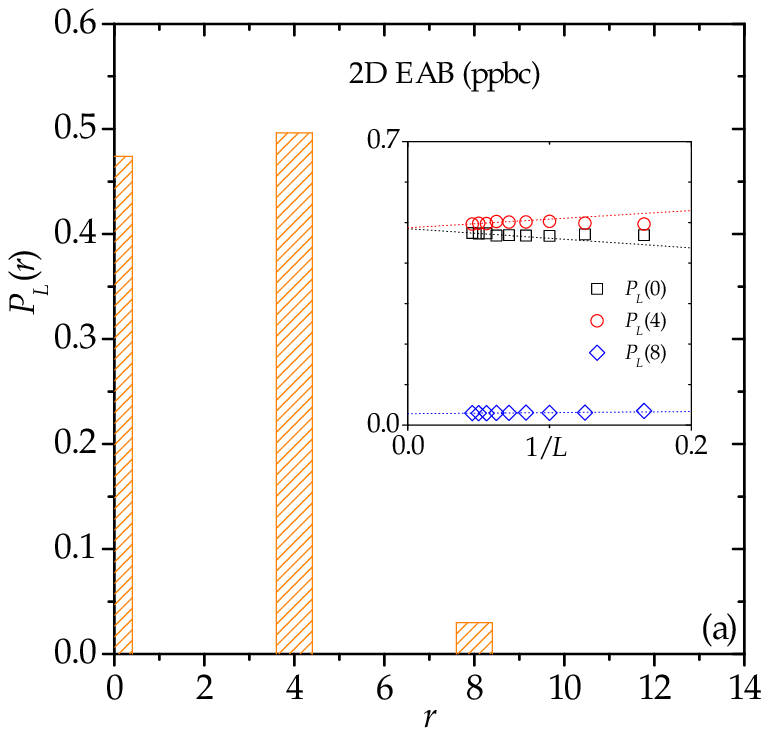}
\includegraphics[width=7cm,clip=true]{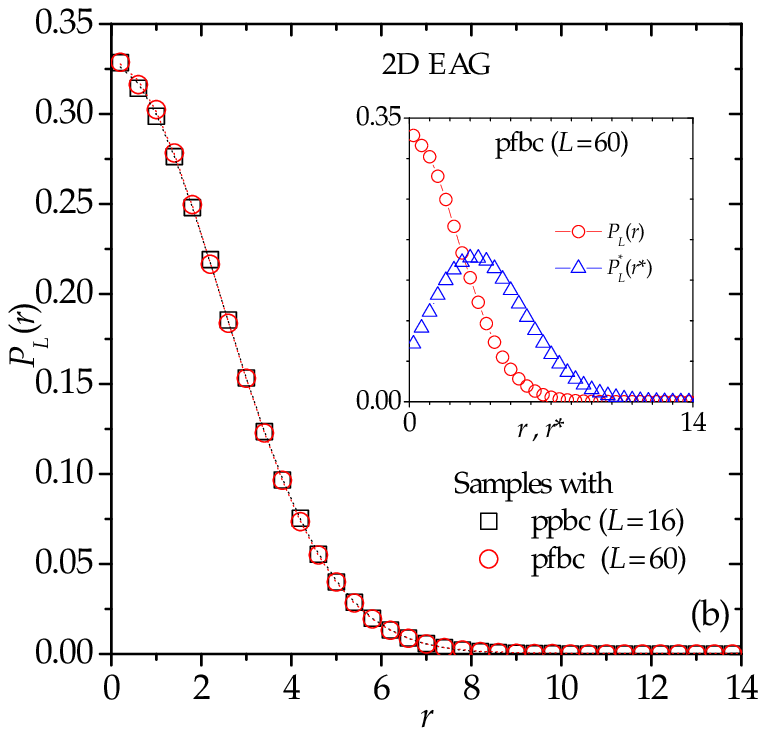}
\caption{\label{figure1} (Color online) Rigidity distributions for the 2D EAB and EAG models model. 
(a) EAB model with ppbc ($L=22$). The inset shows the height of each bar as a function of $1/L$. 
(b) 2D EAG model, with ppbc ($L=16$) and  pfbc ($L=60$). The curves correspond to Gaussian fits 
[see Eq.(\ref{FunGauss})].  The inset shows a comparison between $P_L(r)$ and $P^*_L(r^*)$ for samples with pfbc.}
\end{figure}

We begin by analyzing the rigidity distribution, $P_L(r)$, for 2D models.   Figure \ref{figure1} (a) 
shows this distribution for the 2D EAB model with ppbc and lattice size $L=22$.  Bonds with rigidity 
$r=0$ are flexible bonds while the remaining ones form the rigid lattice.  For lattices with ppbc, the 
only possible non-zero rigidity values are $r=4$ and $r=8$, i.e., the energy difference between the GS 
and the first and second excited states, respectively.  By extrapolating toward the thermodynamic limit, 
we obtain the asymptotic rigidity distribution, given by the following: $P(0)=0.48(1)$, $P(4)=0.49(1)$ and $P(8)=0.028(1)$ 
[see the inset in Fig.\ref{figure1}(a)]. For samples with with pfbc we obtain similar, but smaller, values 
because the presence of the free boundary generates two additional bars at $r=2$ and $r=6$. Even though 
these bars must vanish in the thermodynamic limit, for $L=60$ they are still significant (comprising approximately 
$8\%$ of the bonds), which is consistent with the large finite size effects found for other quantities 
in this system \cite{Roma2010b,Hartmann2001}.

We turn now to the 2D EAG model.  Figure \ref{figure1}(b) shows a comparison between the rigidity distributions 
for systems with ppbc ($L=16$) and with pfbc ($L=60$). Note that finite size effects are not relevant in this 
case. $P_L(r)$ is a continuous function taking appreciable values within a similar range of rigidity as 
for the 2D EAB model. 

Interestingly, the distributions obtained differ substantially from what is obtained by considering 
locally defined quantities.  For each bond $J_{ij}$ of a given sample, we define its {\em local rigidity} 
as $r^*_{ij}=U^*_{ij}-U$, where $U$ is the GS energy and $U^*_{ij}$ is the smallest of the energies of 
the two configurations obtained by flipping either the spin $\sigma_{i}$ or the spin $\sigma_{j}$.  
The inset of Fig.\ref{figure1}(b) shows, for the 2D EAG model, that the distribution of rigidity $P_L(r)$ 
and the distribution of local rigidity, $P_L^*(r^*)$, differ markedly.  The local rigidity is closely 
related to the {\em local fields}~\cite{Marshall1960,Klein1963}: The rigidity of a bond is simply the minimum 
between the absolute local fields, at $T=0$, of spins $i$ and $j$.  There is also a factor $2$ that arises 
from the fact that the rigidity is calculated as an energy difference.  The distribution of local fields 
at $T=0$ has been calculated for the EAG in 2D and 3D, and the curves are very similar to the distribution 
of local rigidities \cite{Boettcher2008}.  The rigidity distribution $P_L(r)$ [Fig.\ref{figure1}(b)] 
can be well fitted by a Gaussian function,  
\begin{equation}
Q(x)=\frac{\sqrt{8/\pi}}{w} \exp \left(-\frac{2 x^2}{w^2}\right), \label{FunGauss}
\end{equation}
where $w$ is a constant.  For example, for the 2D EAG with ppbc we obtain $w=4.89(1)$.  

\begin{figure}[t]
\includegraphics[width=6.5cm,clip=true]{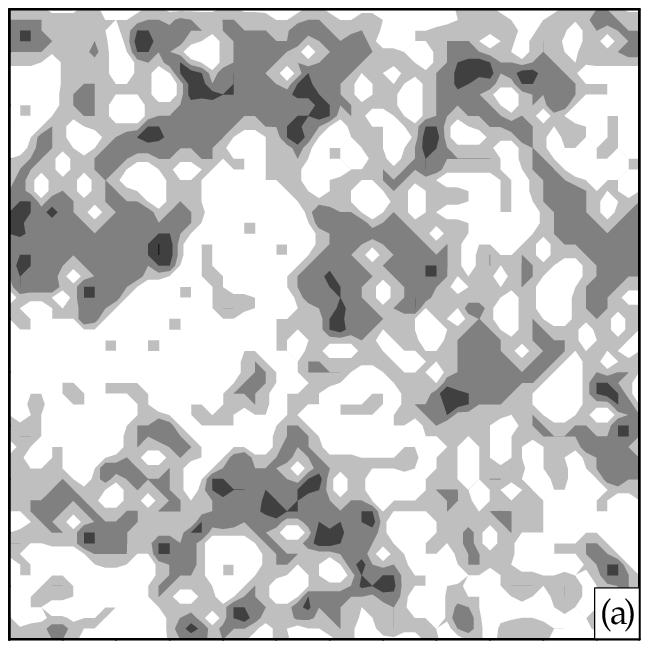}
\includegraphics[width=6.5cm,clip=true]{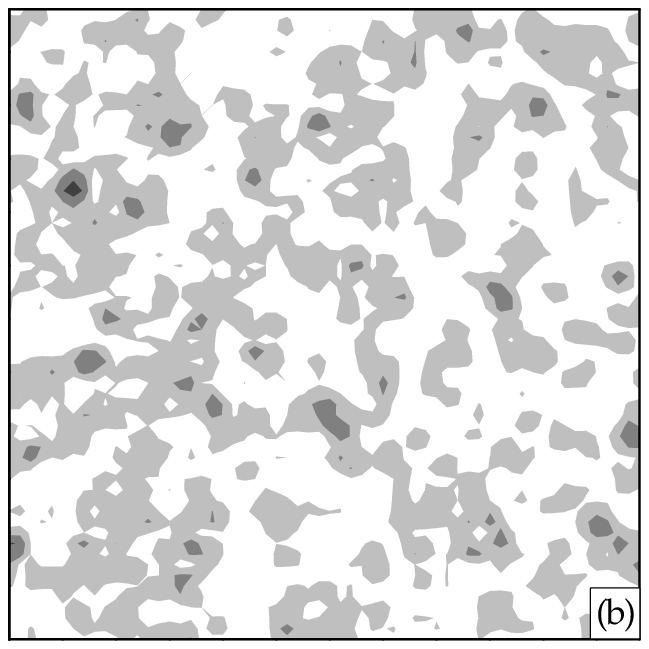}
\caption{\label{figure2} Map plot of the ``average rigidity lattice'' for two samples of the 2D (a) EAB and (b) 
EAG models of $L=60$.  In both figures the grayscale is the same and the average rigidity values are 0 (white), 
2, 4, 6 and 8 (black). }
\end{figure}

Another important feature of the RS is its spatial distribution. Figure \ref{figure2} (a) shows a map plot 
representing the RS of a 2D sample of the EAB model with pfbc and linear size $L=60$.  This map plot was generated 
from an ``average rigidity lattice,'' where the shade of gray at each site value is given by the average rigidity $\bar{r}$ of the four bonds 
connecting with this site.  It can be observed that bonds with similar rigidity are segregated.  Figure \ref{figure2} (b) shows that this occurs also for a 2D 
sample of the EAG model and that there are small differences between their spatial 
rigidity distributions.   

\begin{figure}[t]
\includegraphics[width=7cm,clip=true]{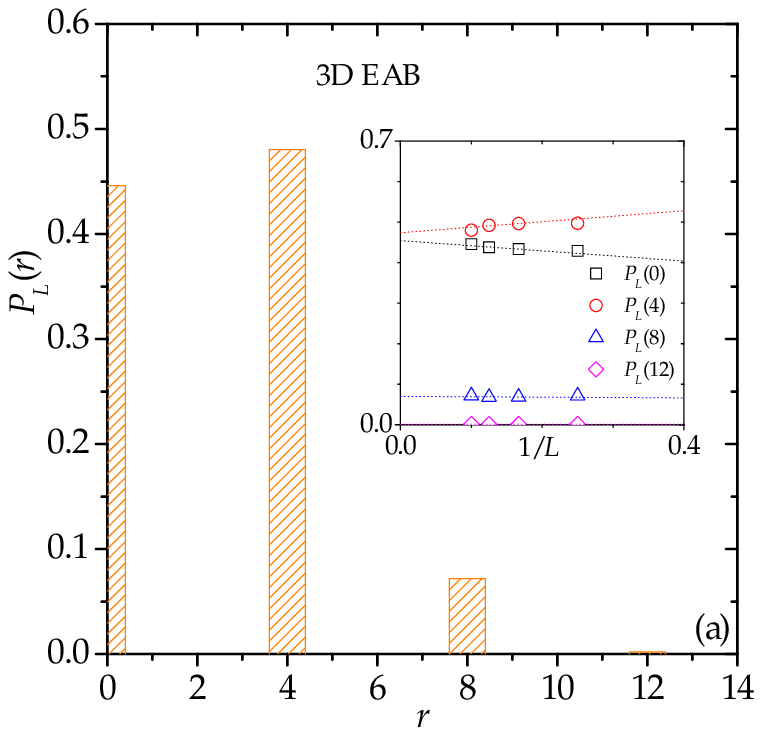}
\includegraphics[width=7cm,clip=true]{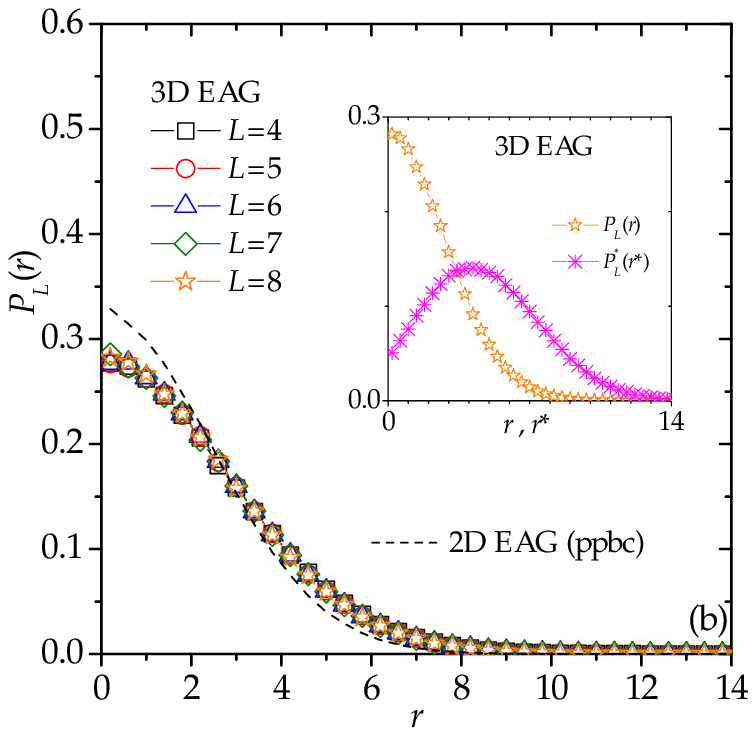}
\caption{\label{figure3} (Color online) (a) Rigidity distribution for the 3D EAB model ($L=10$). The inset 
shows the peaks height of the distribution as function of $1/L$. (b) Rigidity distribution for the 3D EAG model 
for different lattice sizes as indicated, and for the 2D EAG model (samples of size $L=16$ with ppbc). The 
inset shows a comparison between $P(r)$ and $P^*(r^*)$ for 3D samples. }
\end{figure}

Next, we analyze the rigidity distribution $P_L(r)$ for the 3D models.  Figure \ref{figure3} (a) shows this distribution 
for samples of the EAB model of size $L=10$. Now the only possible non-zero rigidity values are $r=4$, 
$r=8$ and $r=12$, which form the rigid lattice.  By extrapolating toward the thermodynamic limit we obtain the 
following asymptotic distribution: $P(0)=0.45(3)$, $P(4)=0.47(2)$, $P(8)=0.07(1)$ and $P(12)=0.0016(4)$ [see the inset in Fig.\ref{figure3}(a)].  
Figure \ref{figure3}(b) shows the distribution $P_L(r)$ for different lattice sizes of the 3D 
EAG model. It can again be seen that the shape of $P_L(r)$ is unaffected by finite-size effects and that the distribution 
for $L=8$ can be well fitted by Eq.(\ref{FunGauss}) with $w=5.61(2)$.  A comparison with 
the rigidity distribution of the 2D EAG model shows that the range of nonvanishing values of $r$ in 3D is slightly 
larger, in agreement with our observations for the 3D EAB model. By comparing $P_L(r)$ with the distribution of 
local rigidity $P_L^*(r^*)$ [the inset in Fig.\ref{figure3}(b)], it is once again clear that both functions differ substantially.  
In addition, by means of a qualitative analysis of the rigidity structure, we have observed that a rigidity segregation also 
emerges in 3D models.  

\subsection{Percolation of the RS in the EAB model}
 
In this and in the next subsection, we discuss how to use the RS to determine which part of the system 
has the relevant properties to be called a backbone.  But first we need to address the question of which 
properties are to be considered relevant. We assume that there is a close connection between the topological 
features of the sets of bonds with a similar rigidity and the equilibrium critical behavior of a given 
spin-glass model. The main conjecture, proposed in Ref. \onlinecite{Roma2010b}, is that, in a system with 
quenched disorder, it is the rigidity $r_{ij}$, and not the bond strength $J_{ij}$, the quantity that gives 
the magnitude of the ``effective interaction'' between spins $i$ and $j$.  If this is true, then the backbone 
of a spin-glass model with a finite (zero) critical temperature should have a percolation cluster with a 
finite (zero) rigidity value.  In addition, within this cluster the correlation length should diverge at 
the critical point. To address this issue, in the following we analyze the percolation properties of sets 
with similar rigidities.

It is very difficult to make a direct study of percolation of the different sets in the EA models, because 
the systems whose RS can be obtained are too small \cite{Roma2010b}. To overcome this problem, we follow 
here a more complex, but more conclusive, approach \cite{Vogel1998a,Vogel1998b,Roma2010b}. In a nutshell, 
what we do, for each structure to be analyzed, is to build a curve of percolation probabilities and to 
extract a percolation threshold from it.  Then we compare this threshold with the estimated size of the 
corresponding structure in the thermodynamic limit.  If this last number is larger than the threshold, we 
conclude that the structure percolates in the thermodynamic limit. In the following the procedure is explained 
in more detail for the EAB model, where the rigidity can only take discrete values.

\begin{figure}[t]
\includegraphics[width=7cm,clip=true]{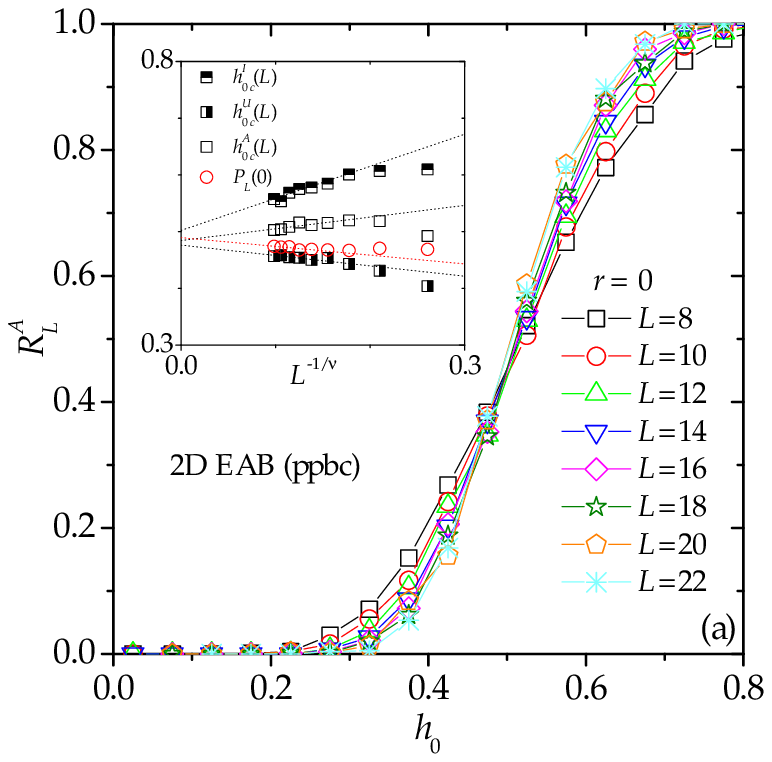}
\includegraphics[width=7cm,clip=true]{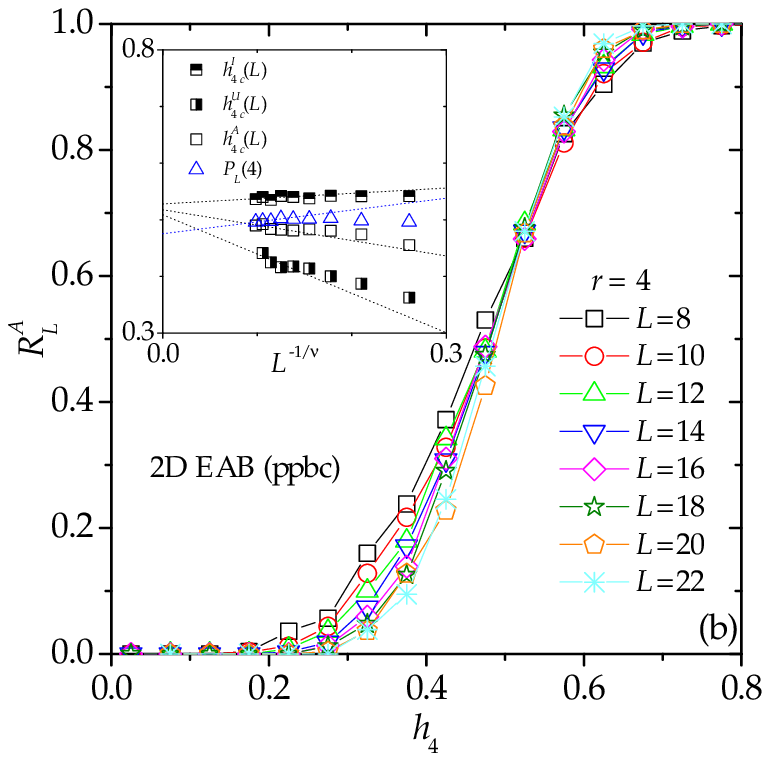}
\caption{\label{figure4} (Color online) Percolation probability $R_L^A$ for the 2D EAB model (ppbc), for the 
substructures of the RS with (a) $r=0$ and (b) $r=4$. The insets show $P_L(r)$ and the effective thresholds 
for the percolation criteria $I$ and $A$, as a function of $L^{-1/\nu}$ for (a) $r=0$ and (b) $r=4$.  }
\end{figure}

For the set of bonds having the same rigidity $r$, we define $R^U_L(h_r)$ and $R^I_L(h_r)$ \cite{Yonezawa1989} 
as the probabilities that this set percolates along at least one lattice direction, and simultaneously along 
all independent lattice directions, respectively, if the size of the set is between $h_r$ and $h_r + \Delta h$ 
(sizes are given as fractions over the total number of bonds). We also define the arithmetic mean of these 
quantities as $R^A_L(h_r) \equiv [R^U_L(h_r)+R^I_L(h_r)]/2 $. 

Figures~\ref{figure4}(a) and \ref{figure4}(b) show, for the 2D EAB model with ppbc, the percolation probability function 
$R^A_L $ of the substructures with $r=0$ and $r=4$, respectively, for different lattice sizes (we show only the percolation criterion $A$ because for this model it is the quantity less sensitive to finite 
size effects). Calculations were performed using the algorithm of Hoshen-Kopelman \cite{Hoshen1976}.  As for 
small lattice sizes the fractions of rigid bonds such as $h_0$ and $h_4$ have a very wide distribution 
\cite{Roma2010b}, the curves in these figures extend over almost the entire range of this variable, from 0 to 1.  In all cases, 
we have used a bin width of $\Delta h = 0.05$. Error bars were calculated using a bootstrap method \cite{numerical} 
but they are omitted when they are smaller than the symbol size.  

Although the crossing points that define the percolation thresholds are easy to establish, we perform for each 
set a standard analysis of the data, in order to improve the accuracy of the values obtained \cite{Yonezawa1989,Roma2010b}. 
First, each curve is fitted with an error function using a least-mean-squares method. Then the concentration 
at which the slope of the fitting curve is largest is taken as an effective threshold $h^\mathcal{X}_{rc} (L)$, 
where $\mathcal{X}$ denotes the percolation criterion used: $U$, $I$ or $A$. $h^\mathcal{X}_{rc} (L)$ is expected 
to follow the law \cite{Stauffer1985}
\begin{equation}
h^\mathcal{X}_{rc}(L)=h_{rc}+C^\mathcal{X} L^{-1 /\nu},
\end{equation}
where $C^\mathcal{X}$ is a non-universal constant and $\nu$ is the critical exponent associated to the correlation 
length. As in a previous work \cite{Roma2010b}, for the 2D EAB model we have used $\nu = 4/3$, corresponding to 2D 
random percolation \cite{Stauffer1985}. 

The inset in Fig.~\ref{figure4}(a) shows the mean fraction of bonds with $r=0$, $P_L(0)$, and the effective thresholds 
for the three percolation criteria.  To calculate an estimate of the percolation thresholds at the thermodynamic 
limit, we have extrapolated the data by means of a linear fit.  We obtain the following
limits: $P(0)= 0.49(1)$, $h_{0 c}^I = 0.506(6)$, 
$h_{0 c}^U = 0.48(1)$ and $h_{0 c}^A = 0.484(8)$ [we determine a somewhat different limit of $P(0)$ from the one 
calculated above, because here the fit is carried out using $L^{-1/\nu}$].  These values are too close to be useful 
to decide whether the substructure with $r=0$ percolates. Unfortunately, even though larger sizes are available, 
the situation is very similar in the case of the 2D EAB model with pfbc. On the other hand, for the substructure with 
$r=4$, we obtain the following: $P(4)= 0.483(6)$, $h_{4 c}^I = 0.53(1)$, $h_{4 c}^U = 0.51(2)$ and $h_{4 c}^A = 0.52(1)$ [see the inset 
in Fig.~\ref{figure4} (b)].  If the percolation threshold $h_{4 c}^U$ is discarded, because of large finite-size effects, 
the conclusion is that this substructure has not percolated. The structure with $r=8$ has not been analyzed 
because its size is too small.          

For 3D lattices we have used $\nu = 0.9$, and only the percolation criterion I because the others are affected by 
large finite size effects.  In the thermodynamic limit we obtain the following: $P(0)= 0.45(1) > h_{0 c}^I = 0.24(1)$ and 
$P(4)= 0.48(2) > h_{4 c}^I = 0.30(1)$.  Thus, unlike what happens in 2D, in 3D both sets with $r=0$ and $r=4$ 
percolate.  The simultaneous percolation of two different structures in 3D systems has also been found in other 
contexts \cite{Isichenko1992}.  

It is interesting to note that the percolation properties of the set of bonds with $r=4$ is very similar to what 
has been found for the whole rigid lattice \cite{Roma2010b}, which comprises all bonds with $r \ge 4$.  
This is because bonds with $r=8$ in 2D, and with $r=8$ and $r=12$ in 3D, can only form small and 
compact clusters that fill the interstices of the larger substructure with $r=4$.

\begin{figure}[t!]
\includegraphics[width=7cm,clip=true]{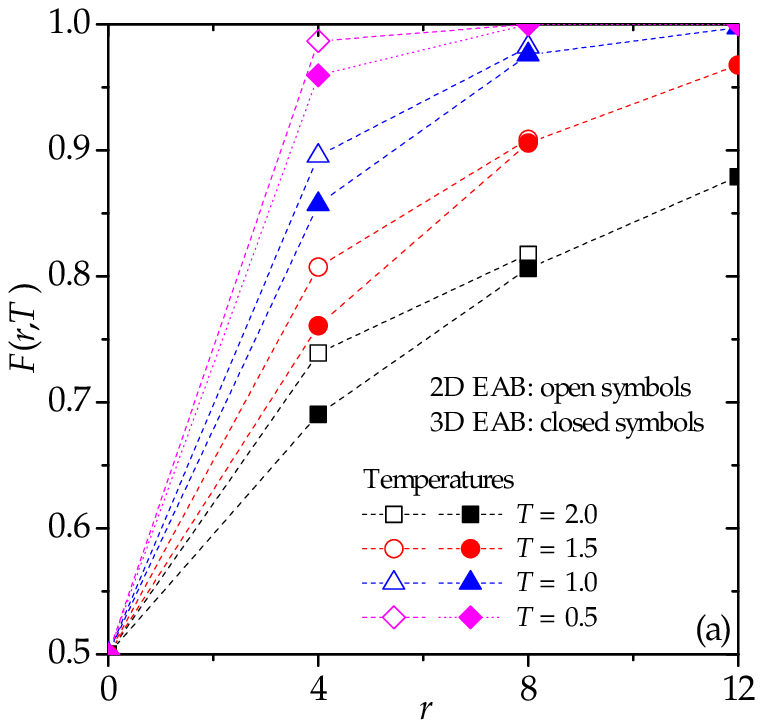}
\includegraphics[width=7cm,clip=true]{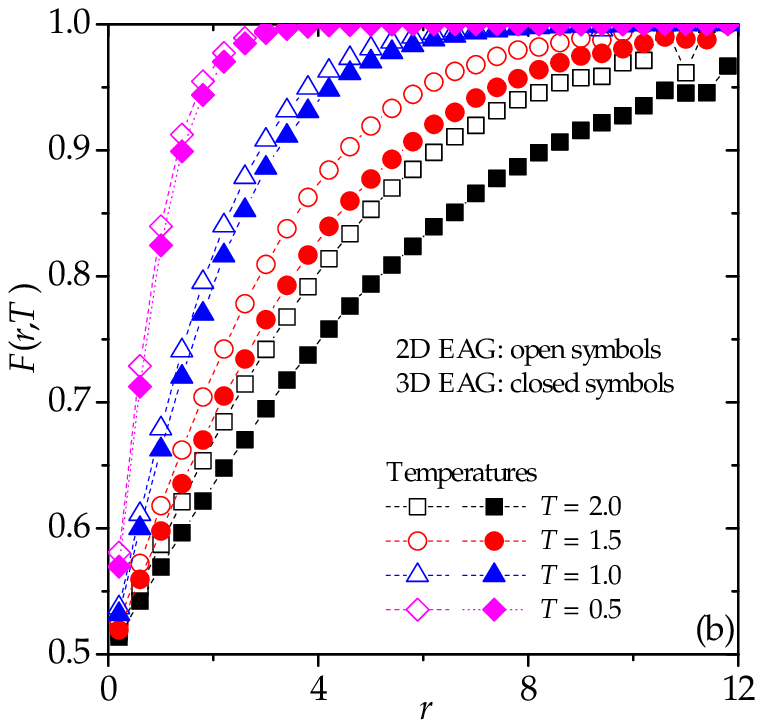}
\caption{\label{figure5} (Color online) $F(r,T)$ as function of $r$ for (a) the 2D and the 3D EAB models and for (b) 
the 2D and the 3D EAG models.  Curves are given for different temperatures as indicated.}
\end{figure}

The idea that the structure with $r=4$ dominates the physical behavior of the backbone is supported by the analysis 
of the function $F_J(r,T)$, defined as the mean value of the fraction of bonds with rigidity $r$ which, at temperature 
$T$, have the same condition (satisfied or frustrated) as in the GS. It is easy to see that, when $T \to 0$, those bonds 
with non zero rigidity frozen and then $F_J(r,T) \to 1$, while for $T \to \infty$ the mean fraction $F_J(r,T) \to 1/2$. 
Figure~\ref{figure5}(a) shows the sample average of this fraction, $F(r,T)$, for the 2D ($L=16$ with ppbc) and the 3D 
($L=8$) EAB models [as bonds with $r=0$ do not have a defined condition on the GS, we have set $F(0,T)=1/2$ for any $T$] 
\cite{Note2}. With decreasing temperature, Fig.~\ref{figure5}(a) shows that bonds with large rigidity freeze faster than 
those with small rigidity.  This gives support to our conjecture that it is the rigidity $r_{ij}$, and not the bond 
strength $J_{ij}$, that gives the magnitude of the effective interaction between spins $i$ and $j$ (recall that in the 
EAB we have $|J_{ij}|=1$).  In particular for the 3D EAB model and close to the critical temperature $T_c \approx 1.12$ 
\cite{Katzgraber2006}, we see that bonds with $r=12$ and $8$ are almost completely frozen, while only the bonds with $r=4$ 
are affected by thermal fluctuations.  But it is precisely this substructure that percolates in the thermodynamic 
limit.  Thus, it is reasonable to assume that, at the critical point, it is in this region that there is a divergence 
of the correlation length.

Figure~\ref{figure5}(b) shows the same function $F(r,T)$ for the 2D and the 3D EAG models, for lattice sizes of $L=16$ 
and $L=8$, respectively.  We observe a similar behavior to that of the EAB, but now for models with continuous bond and 
rigidity distributions.  Notice that close to $T_c \approx 0.95$ \cite{Katzgraber2006}, bonds with a rigidity between 
$r\approx 2$ and $r\approx 4$ are affected by thermal fluctuations.                  

Remarkably, $F(r,T)$ is very well fitted by the functional form $1/(1+\exp(-ar/T))$, where the parameter $a$ is close 
to $2$ for all models.  This represents the probability that a system with only two levels, with an energy difference of $a \, r$,
is in the GS.   We emphasize 
that this result is not a trivial consequence of the definition of bond rigidity, because the $r$ values are a measure 
of the energy difference between the GS and the low-excitation levels (where entropic effects are not taken into account), 
whereas the function $F(r,T)$ is calculated in equilibrium, where the dynamics takes place at finite temperature and the system explores configurations corresponding to highly excited levels.              

\subsection{Percolation of the RS in the EAG model}

To study percolation in the EAG model, we choose to break the RS into two sets, one formed by the bonds with rigidity 
$r \ge r_{\textrm{min}}$, $\Omega(r_{\textrm{min}})$, and another set $\Omega^*(r_{\textrm{min}})$ comprising the remaining 
bonds, where $r_{\textrm{min}}$ is a given rigidity value.  This choice is motivated by the existence of the rigid and the
flexible lattices in EAB systems. Besides, the study of the percolation of smaller structures (centered around given values 
of $r$) is not possible in our case because the size of the systems used is too small.The sizes of the sets 
$\Omega(r_{\textrm{min}})$ and $\Omega^*(r_{\textrm{min}})$, denoted as $h_x$ and $h_x^*=1-h_x$, respectively, have large 
variations for different samples. However, in the previous section we have shown that the rigidity distribution $P_L(r)$ shows 
almost no dependence on sample size for the EAG, and thus the same happens for the sample averages of $h_x$, denoted as 
$X(r_{\textrm{min}})$. 

\begin{figure}[t]
\includegraphics[width=7cm,clip=true]{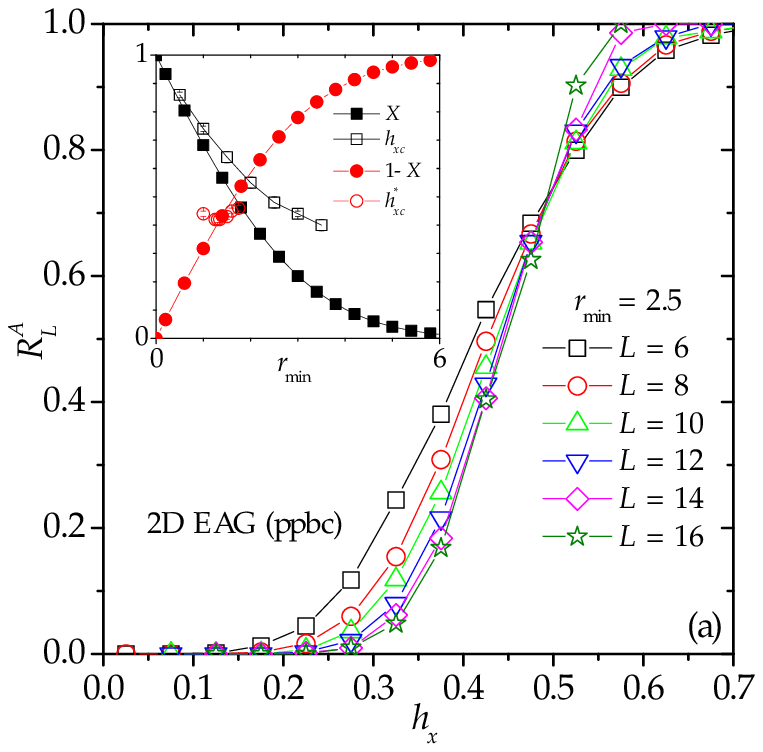}
\includegraphics[width=7cm,clip=true]{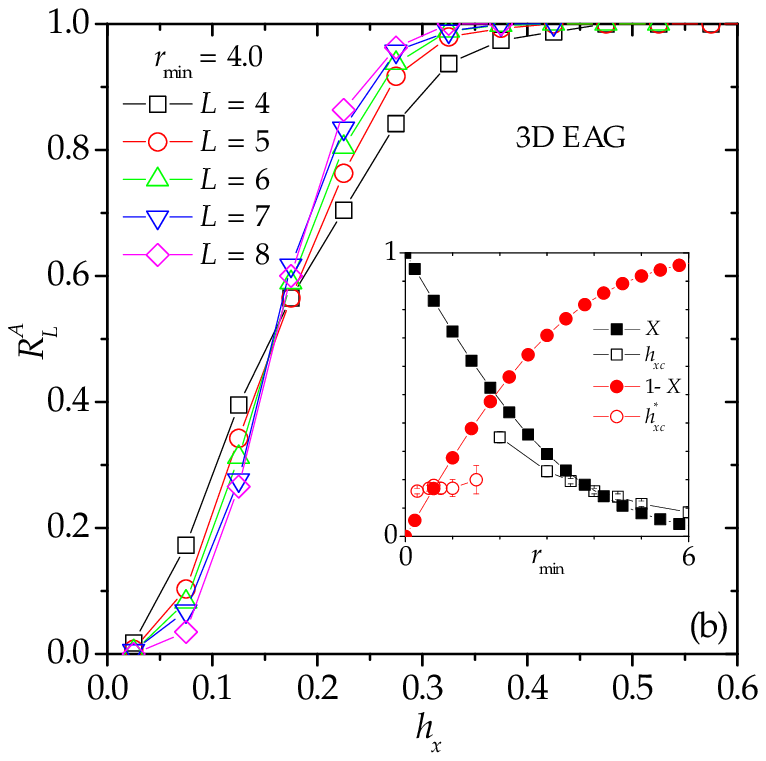}
\caption{\label{figure6} (Color online) Percolation probability $R_L^A$ curves for (a) the 2D (ppbc) and (b) the 
3D EAG models for $r_{\textrm{min}}=2.5$ and $r_{\textrm{min}}=4$, respectively.  The insets show for each model, 
the comparison between $h_{xc}$ and $X$, and between $h^*_{xc}$ and $1-X$, as functions of $r_{\textrm{min}}$.}
\end{figure} 

Similarly to what we have done for the EAB models, for each value ​​of $r_{\textrm{min}}$ we calculate a percolation 
threshold $h_{xc}(r_{\textrm{min}})$ and then we compare this number to the corresponding $X(r_{\textrm{min}})$: If 
$h_{xc}<X$ the set $\Omega$ percolates. Figure \ref{figure6}(a) shows the curves of the percolation probability 
$R_L^A$ for the 2D EAG model with ppbc for $r_{\textrm{min}}=2.5$.  In this case the percolation threshold is 
(calculated as in the previous section) $h_{xc}(2.5)=0.48(2)$, and $X(2.5)=0.31(1)$; therefore ,the set $\Omega(2.5)$ 
does not percolate. For too-small or too-large values of $r_{\textrm{min}}$ the calculation of the threshold 
becomes impossible because of the small size of the corresponding sets in the samples available to us.

The inset in Fig.~\ref{figure6}(a) shows the comparison between $h_{xc}$ and $X$ as function of $r_{\textrm{min}}$.  
Extrapolating we can deduce that, for the 2D EAG model with ppbc, the set $\Omega(r_{\textrm{min}})$ percolates when 
$r_{\textrm{min}} \to 0$, which is consistent with a vanishing $T_c$ for this model (this is similar to what happens 
for the 2D EAB model).                

Now we consider the complementary set $\Omega^*(r_{\textrm{min}})$ composed by the bonds with $r < r_{\textrm{min}}$.  
For this set the percolation threshold is denoted as $h^*_{xc}(r_{\textrm{min}})$.  Comparing this quantity 
with the fraction $1-X(r_{\textrm{min}})$, we can find whether the bonds with the lowest rigidity percolate in the 
thermodynamic limit.  The inset in Fig.~\ref{figure6} (a) shows this comparison.  As we can see for the 2D EAG model, 
when $r_{\textrm{min}} \gtrsim 1.3$ the set $\Omega^*(r_{\textrm{min}})$ percolates.      

For the 3D EAG model the situation differs.  Figure~\ref{figure6}(b) shows that, for the set 
$\Omega(r_{\textrm{min}})$, the percolation probability $R_L^A$ curves for $r_{\textrm{min}}=4$ intersect at 
$h_{xc}(4)=0.16(1)$, a value very close to $X(4)=0.163(1)$. In turn, the inset shows that $h_{xc}$ and 
$X$ cross at a value of $r_{\textrm{min}}$ close to 4.  On the other hand, for the complementary set $\Omega^*(r_{\textrm{min}})$, the inset in Fig.~\ref{figure6}(b) also shows that the curves of $h^*_{xc}$ and $1-X$ cross 
at $r_{\textrm{min}} \approx 0.6$. 
Thus, for $0.16<r_{min}<4$ both sets, $\Omega(r_{\textrm{min}})$ and $\Omega^*(r_{\textrm{min}})$, percolate.

\subsection{The backbone of the EAG models}

\begin{figure}[t]
\includegraphics[width=7cm,clip=true]{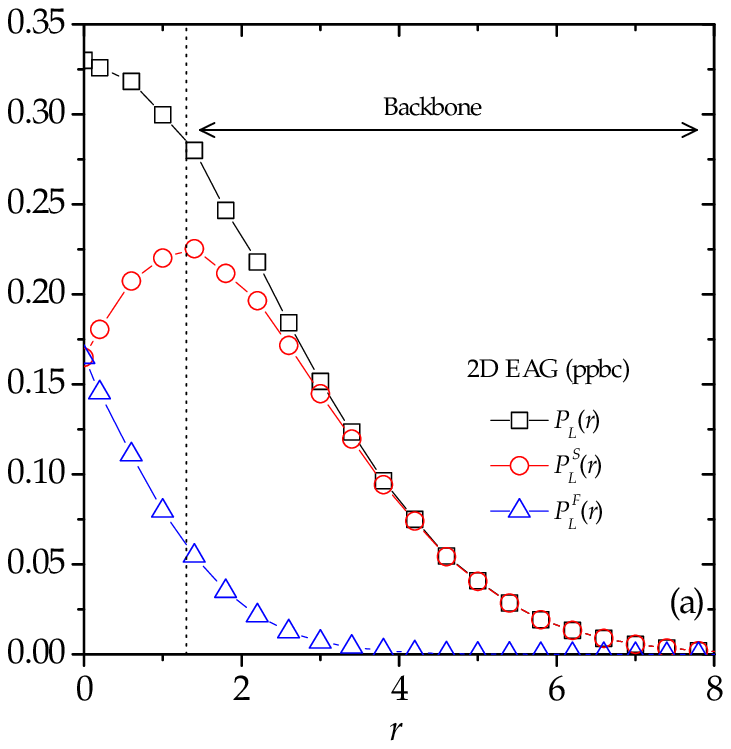}
\includegraphics[width=7cm,clip=true]{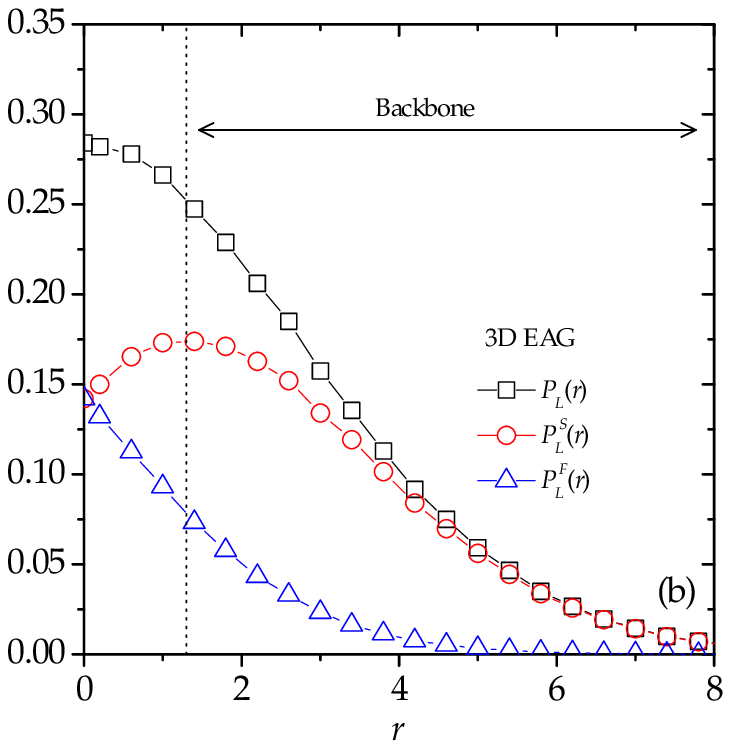}
\caption{\label{figure7} (Color online) Rigidity distributions $P_L(r)$, $P_L^S(r)$, and $P_L^F(r)$ (see text) for 
(a) the 2D  ($L=16$ with ppbc) and (b) the 3D ($L=8$) EAG models.  The vertical dotted lines marks the value of 
$r_{\textrm{min}} = 1.3$.}
\end{figure}

In the previous sections we have shown that the percolation properties of the RS for the EAG and the EAB models are 
very similar. In the following we show that there are values of $r_{\textrm{min}}$ for which the physical properties 
of the sets $\Omega(r_{\textrm{min}})$ and $\Omega^*(r_{\textrm{min}})$ are equivalent, respectively, to those of the backbone and its complement in EAB models. 

Figures~\ref{figure7}(a) and \ref{figure7}(b) show, for the 2D and the 3D EAG models, a comparison between the rigidity distributions 
for all bonds, $P_L(r)$, for the bonds that are satisfied in the GS, $P_L^S(r)$, and for the bonds that are frustrated 
in the GS, $P_L^F(r)$.  Obviously the distributions must satisfy $P_L(r)=P_L^S(r)+P_L^F(r)$.  Note that the distributions 
$P_L^S(r)$ and $P_L^F(r)$ are not normalized to unity because their integrals are equal to, respectively, the fractions of 
satisfied and frustrated bonds of the GS.  In both cases, for large values of $r_{\textrm{min}}$, almost all the bonds in 
the set $\Omega(r_{\textrm{min}})$ are satisfied in the GS.  However, even though with decreasing $r_{\textrm{min}}$ the 
size of the set $\Omega(r_{\textrm{min}})$ increases, the fraction of frustrated bonds is rather small. A change of trend 
occurs when the $P_L^S(r)$ distribution reaches its maximum: For smaller values of $r_{\textrm{min}}$ the set $\Omega(r_{\textrm{min}})$ 
begins to incorporate highly frustrated regions. Remarkably, this maximum happens approximately at $r_{\textrm{min}} \approx1.3$ 
for both, the 2D and the 3D EAG models.

Interestingly, for the 2D EAG model we have shown in the previous subsection that at $r_{\textrm{min}} = 1.3$ 
the complementary set $\Omega^*(1.3)$ is very close to the percolation thresholds, whereas the set $\Omega(1.3)$ does not percolate.    
Therefore, it seems reasonable to define the backbone of this system as the set $\Omega(1.3)$. This choice 
gives a backbone having the $60 \%$ of the bonds of the system and a fraction of frustrated bonds in the GS of only $0.085$.   

\begin{figure}[t!]
\includegraphics[width=7cm,clip=true]{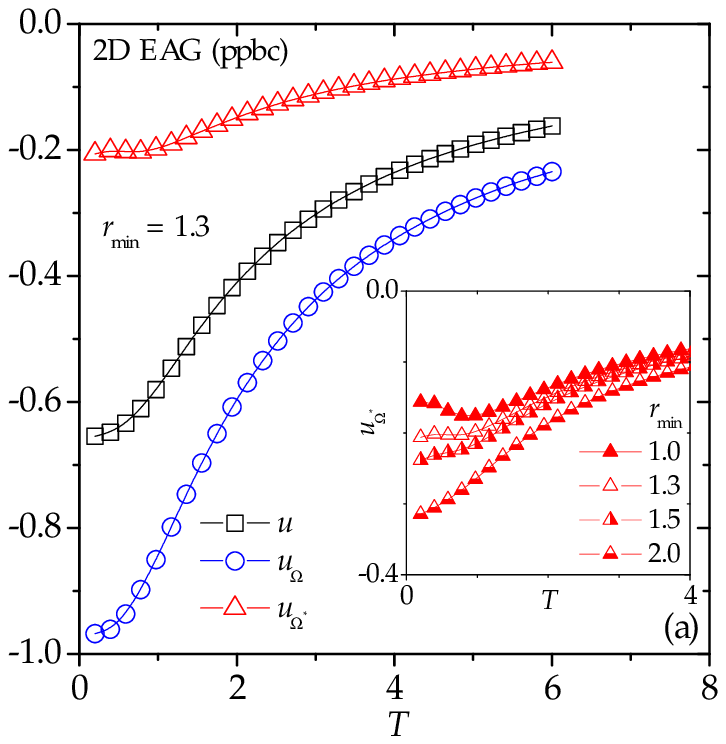}
\includegraphics[width=7cm,clip=true]{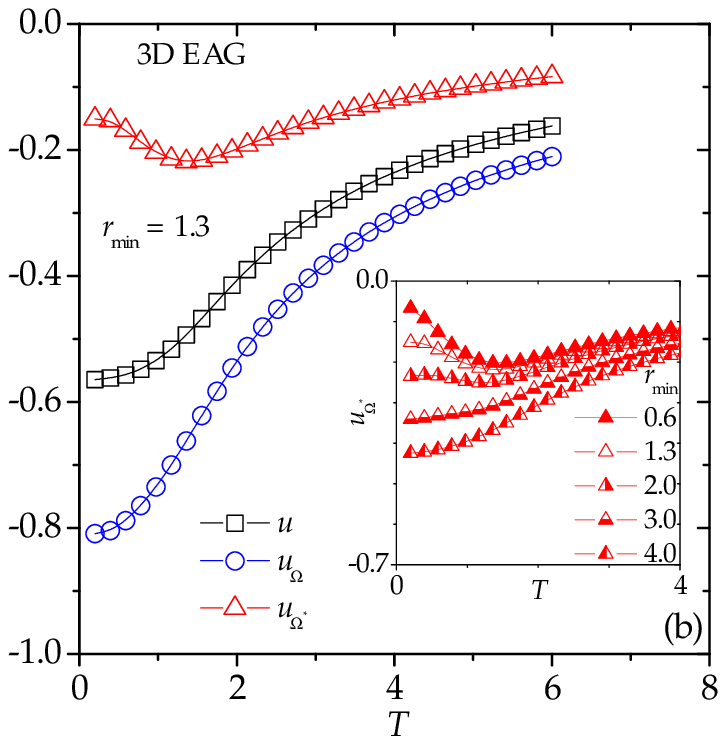}
\caption{\label{figure8} (Color online) Average energies per bond $u(T)$, $u_\Omega(T)$ and $u_{\Omega^*}(T)$ (see text) 
for (a) the 2D  ($L=16$ with ppbc) and (b) the 3D ($L=8$) EAG models.  In both cases $r_{\textrm{min}} = 1.3$ was chosen.  
Insets show the curve of $u_{\Omega^*}(T)$ for different values of $r_{\textrm{min}}$. }
\end{figure}

In the EAB model, several observables behave differently when they are calculated within or outside of the backbone,  
such as the average energy per bond \cite{Roma2010b}.  For example, when this quantity is evaluated outside the backbone, 
it has a minimum at low temperature.  To perform a similar analysis for the EAG models, we define $u(T)$, $u_\Omega(T)$, 
and $u_{\Omega^*}(T)$ as the average energies per bond at temperature $T$ of, respectively, the whole system and the sets 
$\Omega$ and $\Omega^*$ for a given  $r_{\textrm{min}}$. 

Figure \ref{figure8}(a) shows, for the 2D EAG model, these energies as functions of $T$ for $r_{\textrm{min}} = 1.3$.  
Just as in the 2D EAB model \cite{Roma2010b}, there is a very broad minimum in the curve of $u_{\Omega^*}(T)$.  For higher 
(lower) values of $r_{\textrm{min}}$, the inset in that figure shows that the minimum disappears (becomes narrower).  

For the 3D EAG model, Fig.~\ref{figure8}(b) and its inset show that for some values of $r_{\textrm{min}}$ around $1.3$, 
the curves of $u_{\Omega^*}(T)$ display a minimum.  Our calculations suggest that a suitable backbone could be defined 
in the range $[0.6-2.0]$, since all these structures have similar topological characteristics.  However, in the following 
we use $r_{\textrm{min}} = 1.3$ to define the backbone of the 3D EAG model. This set has $64 \%$ of the bonds of the system and a fraction of frustrated bonds in the GS of $0.163$.  

Note that, whereas the percentage of bonds comprising the backbone is only slightly larger in 3D than in 2D, in 3D the 
fraction of those bonds that are frustrated is twice that in 2D.  The same is obtained for the EAB models: The backbone 
(defined as the set of bonds with rigidity $r \ge 4$) comprises approximately $52 \%$ of the system in 2D and $54 \%$ 
in 3D, and the fraction of frustrated bonds is $0.05$ in 2D and $0.1$ in 3D \cite{Roma2010b}.             

\begin{figure}[t!]
\includegraphics[width=7cm,clip=true]{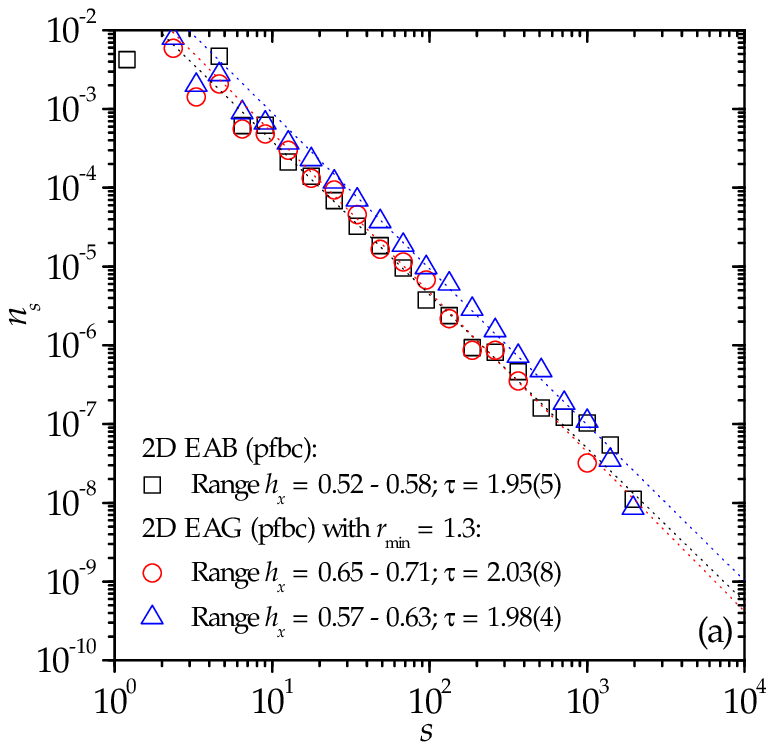}
\includegraphics[width=7cm,clip=true]{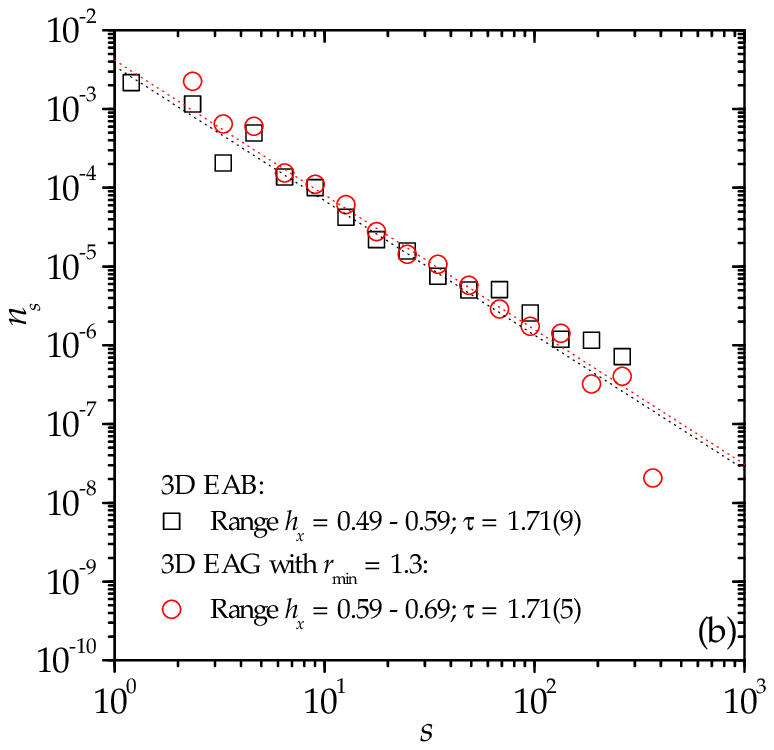}
\caption{\label{figure9} (Color online) Cluster number distributions for the EAB and EAG models for (a) 2D ($L=60$ with pfbc) 
and (b) 3D ($L=8$) lattices.  The distributions are calculated for different ranges.}
\end{figure}

Finally, we analyze the internal structure of the backbone by studying the cluster number distribution, $n_s$, i.e., the number 
of clusters of size $s$.  For the random bond percolation at the critical concentration, it is expected that this distribution 
follows a power law
\begin{equation}
n_s \propto s^{-\tau},
\end{equation}
where $\tau$ is a critical exponent \cite{Stauffer1985}.  Because large samples are needed, in 2D we have
restricted our analysis to lattices with pfbc of size $L=60$.  Figure~\ref{figure9}(a) shows, for the 2D EAB model, 
the cluster number distribution calculated for a range of $h_x$ centered at $0.55$, the mean fraction of bonds with 
rigidity $r \ge 4$.  Fitting the curve we obtain $\tau = 1.95(5)$.  On the other hand, we have also calculated this 
distribution for the backbone of the 2D EAG model ($r_{\textrm{min}} = 1.3$), for two different ranges of $h_x$: one 
centered at $h^*_{xc}(1.3)=0.68$ and another at $X(1.3)=0.6$ [see the inset in Fig.~\ref{figure6} (a)].  For these ranges 
we obtain, respectively, $\tau = 2.03(8)$ and $\tau = 1.98(4)$. Whatever the model or the range, the exponent values 
are very close to $\tau=187/91 \approx 2.05$, the corresponding exponent for the 2D random percolation universality 
class \cite{Stauffer1985}.      

For 3D systems Fig.~\ref{figure9}(b) shows the cluster number distributions for both the 3D EAB model for a range 
centered at $0.54$ (the mean fraction of bonds with rigidity $r \ge 4$), and the 3D EAG model for a range centered at 
$X(1.3)=0.64$ [see the inset in Fig.~\ref{figure6} (b)].   We obtain, respectively, $\tau = 1.71(9)$ and $\tau = 1.71(5)$, 
values that clearly differ from the accepted exponent $\tau = 2.2$ of random percolation in 3D \cite{Stauffer1985}.  
This difference is presumably due to the fact that, unlike the 2D case, here the mean size of the backbone tends to a 
value far from the percolation threshold.  Nevertheless, for the small sizes considered, we see that the cluster number 
distributions for both 3D models follows a power law with the same exponent.  These results show that the internal 
structures of the backbones of systems with bimodal and Gaussian bond distributions are similar.           

\section{Discussion and conclusions \label{Conclusions}}

It has recently been shown that in EA models with a discrete distribution of bonds, and therefore with a degenerate 
ground state, it is possible to find a set of bonds, called backbone, which is closely related to the heterogeneities 
of the GS structure, and that may be used to have a better understanding of the physical properties of such systems. 
More specifically, the backbone and its complement seem to influence both the equilibrium and the out-of-equilibrium 
dynamics of the EAB models \cite{Roma2006,Roma2007b,Roma2010a,Roma2010b,Rubio2010a,Rubio2010b}. Here we have shown 
that in systems with continuous distributions of bonds it is possible to define a continuous version of bond rigidity, 
which in turn leads to the definition of a backbone. Even though there is a certain degree of arbitrariness in the 
choice of the threshold rigidity value that defines the backbone, we have shown here that the resulting structure shares
most of the topological characteristics and physical properties of the backbone of EAB systems.

We have also argued that the reason why sets of bonds with different rigidities (as the backbone and its complement) 
have different physical properties, is that the rigidity provides more physical information than the bond strength. 
It could even be thought that the rigidity $r_{ij}$ in some way encodes the magnitude of the effective interaction 
between spins $\sigma_i$ and $\sigma_j$.  To provide additional evidence that the backbone is not directly related 
to bond strengths, we calculate for the 2D and the 3D EAG models the probability distribution that a bond has strength 
$J$ and rigidity $r$.  Dividing this function by $D_{\mathrm{G}}(J)$, the Gaussian bond distribution, we obtain the 
conditional probability density $W(J,r)$ which, for fixed value of $r$, is normalized to unity.

\begin{figure}[t!]
\includegraphics[width=7cm,clip=true]{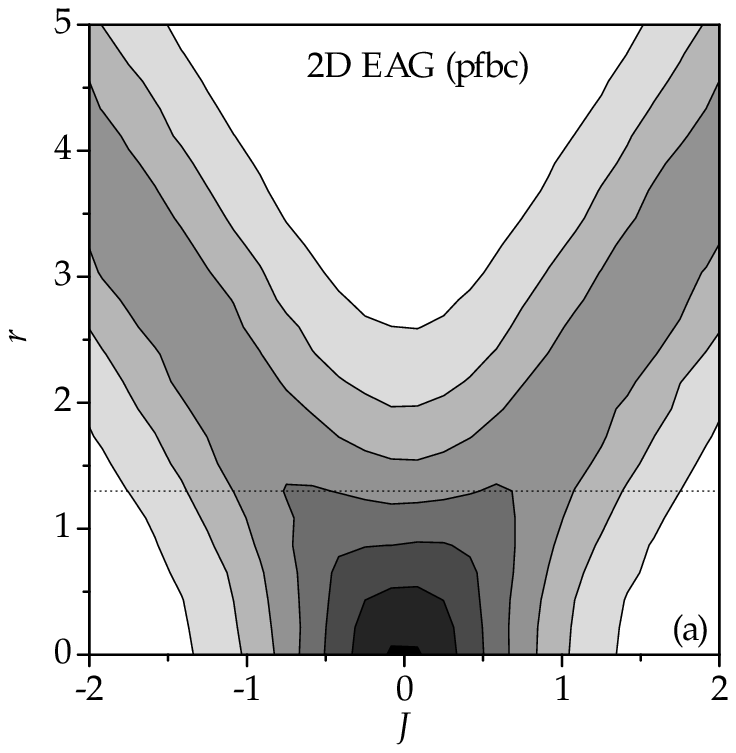}
\includegraphics[width=7cm,clip=true]{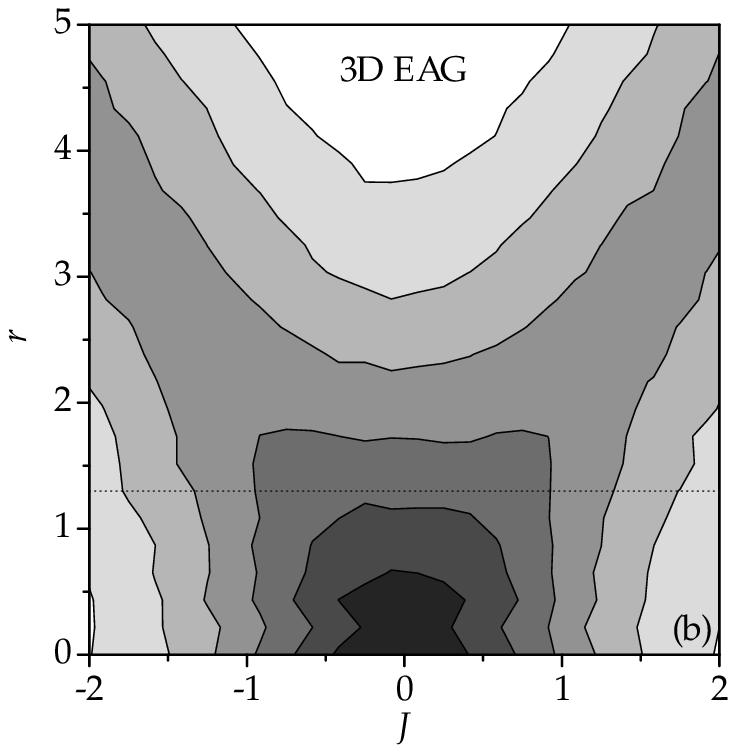}
\caption{\label{figure10} Map plot of the conditional probability density $W(J,r)$ for (a) the 2D ($L=60$ with pfbc) 
and (b) the 3D ($L=8$) EAG models.  The horizontal dotted lines marks the value of $r_{\textrm{min}} = 1.3$.}
\end{figure}

Figures~\ref{figure10}(a) and \ref{figure10}(b) show the map plots of $W(J,r)$ for, respectively, the 2D and the 3D EAG models. 
For relatively large values of $|J|$ and $r$ both distributions develop two arms.  Thus, for most of the bonds in this 
range their strength is proportional to their rigidity.  This is to be expected, since these bonds must be surrounded 
by many others of lower strength, and then the changes in the GS energy (and therefore in the rigidity), produced by 
changing the condition of a bond of great magnitude, should depend primarily on the value of $|J|$. In other words, 
the rigidity of a high-strength bond seems to be, on average, a trivial quantity \cite{Note1}.     

On the other hand, for intermediate and low values of $|J|$ and $r$, the function $W(J,r)$ for both the 2D EAG and 
the 3D EAG models has a square shape in which the approximate proportionality between the strength of a given bond 
and its rigidity is lost. For example in the 3D case, Fig.~\ref{figure10}(b) shows that $W(J,r)$ is almost constant 
on the region given by $-1 \lesssim J \lesssim 1$ and $1 \lesssim r \lesssim 2$.  The same applies to bonds with 
$|J|\approx 1$ which have, with equal probability, a rigidity between $0 \le r \lesssim 2$. These examples show that 
the intermediate region of parameters, where many of the bonds that make up the backbone are located, and where the 
thermal critical (in 3D) fluctuations are important, is nontrivial.   

As mentioned in Sec. \ref{Intro}, temperature chaos \cite{McKay1982,Bray1987,Rizzo2003,Parisi2010} 
is an important issue which must be reconciled with this phenomenological backbone picture. 
If our results were interpreted as a suggestion that ground state excitations can have an influence 
on finite temperature dynamics, the phenomenon of temperature chaos would imply that the backbone can 
only be relevant for the physics of small systems. However, what we argue here is that the system can 
be divided into two macroscopic sectors where the effective interactions between spins, which we assume 
are temperature-independent quantities, seem to differ markedly. The GS and its excitations are only 
used as {\it tools} to find which spins and bonds comprise those sectors. Once the sample has been so divided, 
the information about the states is no longer necessary.  In other words, we do not make any assumptions about 
the magnitude of the overlap between states at finite temperature and the GS and its excitations. Therefore, 
in principle, we would expect that this physical separation of the system is also valid for macroscopic samples. 
However, this conjecture can only be confirmed by performing simulations in those rare small samples 
where temperature chaos has been shown to be present \cite{Katzgraber2007,Fernandez2013}.

The topological characteristics of the backbone may also be relevant for a better understanding of
the physical behavior of other random systems. For example, Tsomokos {\em et al.} \cite{Tsomokos2011}
predict that if the backbone of the 2D EAB model does not percolate, in the random-field toric code model there may
exist a new intermediate quantum phase where topological and spin-glass order coexist.  Our calculations suggest 
that this could also be the case for both 2D models.  In addition, in a recent study of the out-of-equilibrium 
dynamics of the 2D $\pm J$ Potts model at low-temperature, numerical evidence has been found that hints at the 
existence of an underlying backbone structure for this system \cite{Ferrero2012}.  Unfortunately, the RS studied 
there is defined only for models with Ising spins (the $q=2$ case in the $\pm J$ Potts model correspond to the EAB model).  
Although a general procedure to obtain the backbone of an arbitrary system has not yet been found, we believe that 
the progress made here represents a significant step in this direction.  We hope that a further 
generalization of the rigid structure analyzed here will make it possible to identify the backbone of more complex disordered 
models. 

\begin{acknowledgments}

We thank S. Bustingorry, P. M. Gleiser, and L. F. Cugliandolo for fruitful discussions. F. Rom\'a acknowledges financial 
support from CONICET (Argentina) under project PIP 114-201001-00172 and Universidad Nacional de San Luis (Argentina) 
under project PROIPRO 31712.  
\end{acknowledgments}

\end{document}